\documentclass[a4paper,twocolumn]{revtex4}
\usepackage{graphics}
\usepackage{amsmath}
\usepackage{psfig}

\begin{document}

\title{Enhanced Nonlinear Generation in a Three-Level Medium
with Spatially Dependent Coherence}

\author{E. Paspalakis}

\affiliation{Materials Science Department, School of Natural
Sciences, University of Patras, Patras 265 04, Greece}

\author{Z. Kis}

\affiliation{Research Institute for Solid State Physics and
Optics, P.O. Box 49, H-1525 Budapest, Hungary}
\date{\today}

\begin{abstract}
  We consider a method for efficient parametric generation of a laser
  pulse. A single laser field is injected to a three-level medium
  which has two lower states and one excited state. The lower states
  are prepared initially in a position-dependent coherent
  superposition state.  It is shown that by proper choice of the
  position dependence of the superposition along the direction of
  propagation, the incoming field can be converted completely to a new
  field.
\end{abstract}


\maketitle



The problem of resonant nonlinear optics with high efficiency
using phase-coherent media has attracted a lot of interest
recently \cite{Lukinrev}. The usual systems used for this purpose
are the three-level $\Lambda$-type system
\cite{Eberly96a,Harris97b,Harshawardan98a,Deng98a,Vemuri98a}, and
various configurations of multi-level systems
\cite{Lukin98a,Korsunsky99a,Boyd00a,Paspalakis02a}. Some very
interesting experiments using these systems have been performed
\cite{Lukinrev,Hemmer95a,Jain96a,Merriam99a,Merriam00a,Huss01a},
demonstrating the potential for applications of these phenomena.
In a similar area of studies, Grobe and co-workers have shown
that a dielectric medium with initial spatial excitation can have
several novel properties \cite{Csesznegi98a}. In this work, we
study the potential for nonlinear conversion between two laser
pulses using a $\Lambda$-type system with initial spatial
excitation. We show that the existence of a spatially dependent
coherence can lead to parametric generation of a new laser pulse
with unity conversion efficiency.

\begin{figure}[h]
\centerline{\includegraphics{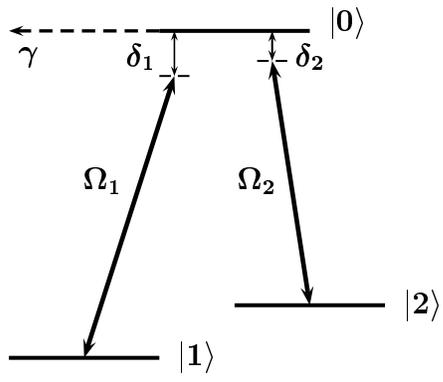}}
\caption{Paspalakis and Kis}
\label{fig1}
\end{figure}

The quantum system under consideration is shown in Fig.\
\ref{fig1}. The spatio--temporal dynamics of the system can be
described by the coupled Maxwell-Schr\"{o}dinger equations in the
rotating wave, dipole, and slowly varying envelope approximations
\begin{subequations}
\begin{eqnarray}
  i\frac{\partial}{\partial \tau} {\boldsymbol b}(\zeta,\tau)
  &=&{\boldsymbol H}(\zeta,\tau){\boldsymbol b}(\zeta,\tau) \; ,
  \label{final1} \\
  \frac{\partial}{\partial \zeta} \Omega_{n}(\zeta,\tau) &=& i
  a_{n}b_{n}(\zeta,\tau)b^{*}_{0}(\zeta,\tau) \, , \label{v0} \quad
  n=1,2 \, ,
\end{eqnarray}
\end{subequations}
with
\begin{eqnarray}
  {\boldsymbol H}(\zeta,\tau)=\left[
    \begin{array}{ccc}
      \delta_{1} -i \frac{\gamma}{2} & \Omega^{*}_{1}(\zeta,\tau) &
      \Omega^{*}_{2}(\zeta,\tau)\\
      \Omega_{1}(\zeta,\tau) & 0 & 0\\
      \Omega_{2}(\zeta,\tau) & 0 & \delta_{1} -  \delta_{2}
    \end{array}
  \right] \, , \label{H1}
\end{eqnarray}
and ${\boldsymbol b}(\zeta,\tau)=\left[b_{0}(\zeta,\tau), b_{1}(\zeta,\tau),
  b_{2}(\zeta,\tau)\right]^{T}$.  These equations are written in the local
frame where $\zeta=z$ and $\tau = t - z/c$. Here, $\Omega_{n}(\zeta,\tau)$ are
the Rabi frequencies and $\delta_{n}$ are the laser field detunings from
resonance. Also, $\gamma$ denotes the decay rate of the excited state out of
the system and $a_{n}$ the propagation constant.

We assume that the system is initially prepared in a superposition of the
lower levels, i.e. we assume a phaseonium medium which was first proposed by
Scully \cite{Scully91a}, such that
\begin{equation} \label{superpos}
  |\psi(\zeta,\tau=0)\rangle = b_{1}(\zeta)|1\rangle + b_{2}(\zeta)|2\rangle
   \, ,
\end{equation}
with $b_{1}(\zeta)$ and $b_{2}(\zeta)$ being, in general, complex
satisfying $|b_{1}(\zeta)|^2+|b_{2}(\zeta)|^2=1$. As it has been
analyzed in detail by Csesznegi et al. \cite{Csesznegi98a} any
desired coherent superposition of such type can be created with
the use of stimulated Raman adiabatic passage (STIRAP)
\cite{Bergmann98a}. This technique uses two laser pulses, ordered
in a counterintuitive sequence, that are applied in the
preparation stage to the medium. The shape of these pulses
\cite{Csesznegi98a} will determine the specific form of
$b_{1}(\zeta)$ and $b_{2}(\zeta)$. We also assume that the
two-photon resonance condition, $\delta_{1} = \delta_{2} =
\delta$, is satisfied. If the excited state $|0\rangle$ decays
rapidly and the laser-matter interaction is weak, so that the
following relations $|\Omega_{n}| \ll \gamma$, $\gamma \bar{\tau}
\gg 1$, $|\Omega_{n}|^{2}\bar{\tau} \ll \gamma$ are satisfied,
with $\bar{\tau}$ being a characteristic pulse length, the
approximate solution of Eq.\ (\ref{final1}) is
\cite{Eberly96a,Paspalakis02a}
\begin{eqnarray}
  b_{1}(\zeta,\tau) &\approx& b_{1}(\zeta) , \quad b_{2}(\zeta,\tau)
  \approx b_{2}(\zeta) \, , \nonumber \\
  b_{0}(\zeta,\tau) &\approx& -   \frac{\Omega^{*}_{1}(\zeta,\tau)b_{1}(\zeta)
  + \Omega^{*}_{2}(\zeta,\tau)b_{2}(\zeta)}{\delta - i \gamma/2} \, .
  \label{b3b}
\end{eqnarray}
and the propagation equation for the laser fields, Eqs.\
(\ref{v0}), reduce to
\begin{equation}
  \frac{\partial}{\partial \zeta} {\boldsymbol \Omega}(\zeta,\tau)
  = -i {\boldsymbol K}(\zeta) {\boldsymbol \Omega}(\zeta,\tau) \, ,
  \label{propag}
\end{equation}
with
\begin{eqnarray}\label{kdef}
  \boldsymbol{ K}(\zeta) = \left[
    \begin{array}{cc}
      \alpha_{1} |b_{1}(\zeta)|^2 & \alpha_{1} b_{1}(\zeta) b^{*}_{2}(\zeta) \\
      \alpha_{2} b_{2}(\zeta) b^{*}_{1}(\zeta) & \alpha_{2}
      |b_{2}(\zeta)|^2
    \end{array}
  \right] \, .
\end{eqnarray}
Here, $\alpha_{n}\!=\! a_{n}/(\delta + i \gamma/2)$ and the vector
of the Rabi frequencies is given by ${\boldsymbol
\Omega}(\zeta,\tau)\! =\! \left[ \Omega_{1}(\zeta,\tau),
\Omega_{2}(\zeta,\tau)\right]^{T}$.

In this work we restrict the discussion to the case when the
propagation constants are equal $a_1=a_2$, which leads to
$\alpha_{1}=\alpha_{2}=\alpha$. However, our results apply also
to the case that $a_{1} \neq a_{2}$, which will be studied in
detail elsewhere. As we are interested in the phenomenon of
parametric generation in our system we will take the initial
conditions as $\Omega_{1}(\zeta=0,\tau)$ = $\Omega(\tau)$ and
$\Omega_{2}(\zeta=0,\tau)$ = 0. In the case that the probability
amplitudes of the phaseonium are independent of $\zeta$, i.e.
$b_{1}(\zeta) = b_{1}$, $b_{2}(\zeta) = b_{2}$, then the solution
of Eq.\ (\ref{propag}) is given by \cite{Eberly96a}
\begin{subequations}
\begin{eqnarray}
\Omega_{1}(\zeta,\tau) &=& \left(|b_{1}|^{2} e^{-i\alpha\zeta} +
   |b_{2}|^{2}\right)\Omega(\tau) \, , \label{v1a2}\\
\Omega_{2}(\zeta,\tau) &=&
 b_{1}^{*}b_{2} \left(-1 +
e^{-i\alpha\zeta}\right)\Omega(\tau) \label{v2a2}\, .
\end{eqnarray}
\end{subequations}
The long distance result, which is mathematically obtained in the
limit $\zeta \rightarrow \infty$ and in practice it means $\zeta
\gg |1/Im(\alpha)|$, is independent of $\alpha$, as
$\Omega_{1}(\zeta \rightarrow \infty,\tau) =
|b_{2}|^{2}\Omega(\tau)$ and $\Omega_{2}(\zeta \rightarrow \infty
,\tau) = -b_{1}^*b_{2}\Omega(\tau)$. This means that complete
conversion is not possible in this case.

The results are very different in the case that $b_{1}(\zeta)$,
$b_{2}(\zeta)$ are not constant. First of all, there is no general
analytic solution of the propagation equation (\ref{propag}). This
equation resembles Schr\" odinger equation with the replacement $\zeta
\leftrightarrow t$, where the propagator $\boldsymbol{K}(\zeta)$, Eq.\
(\ref{kdef}), plays the role of the Hamiltonian and is non-Hermitian
in this case. In the case of a time-dependent Hamiltonian general
solutions can be obtained if the dynamics satisfies adiabaticity
\cite{messiah}.  Therefore, we study the adiabatic evolution of the
system \cite{note}.

The adiabatic evolution is succeeded if the condition
\begin{equation}
  |v(\zeta)|\!\ll\!|\alpha| \,  \label{adiab}
\end{equation}
is fulfilled, with
\begin{equation}
  v(\zeta)\!=\!i\left(b_1(\zeta) \frac{\partial\, b_2(\zeta)}{\partial \zeta}
    - \frac{\partial\, b_1(\zeta)}{\partial \zeta} b_2(\zeta)
  \right) \, .
\end{equation}
These two equations show that in the adiabatic limit the spatial
variation of the probability amplitudes should be smooth and their
rate of change to be smaller than $|\alpha|$. The adiabatic
solutions of Eq.\ (\ref{propag}) is then given by
\begin{subequations}
\begin{eqnarray}
  \Omega_{1}(\zeta,\tau) &=& \left[b_{2}(0)b_{2}(\zeta)^* +
    e^{-i\alpha\zeta}b_{1}(0)^*b_{1}(\zeta)\right]\Omega(\tau)
  \, , \label{v1} \\
  \Omega_{2}(\zeta,\tau) &=& \left[-b_{2}(0)b_{1}(\zeta)^* +
    e^{-i\alpha\zeta}b_{1}(0)^*b_{2}(\zeta)\right]\Omega(\tau) \, .
  \label{v2}
\end{eqnarray}
\end{subequations}
If we choose $b_{1}(0)=0$, $b_{2}(0)=e^{i\phi_2}$ and $b_{1}(\zeta \rightarrow
\infty)=e^{i\phi_1}$, $b_{2}(\zeta \rightarrow \infty)=0$, then in the
adiabatic limit
\begin{eqnarray}
  \Omega_{1}(\zeta \rightarrow \infty,\tau) = 0 \, ,
  \Omega_{2}(\zeta \rightarrow \infty,\tau) = -e^{i(\phi_2 -
    \phi_1)}\Omega(\tau) \, , \label{vfin}
\end{eqnarray}
which means that $|\Omega_{2}(\zeta \rightarrow \infty,\tau)|^2 =
|\Omega(\tau)|^2$. Therefore, by proper choice of the initial
spatial excitation of the system complete nonlinear conversion is
possible between the two laser fields. One may notice that the
choice of the initial distribution is counterintuitive: we need
at the entrance of the medium $b_{1}(\zeta) = 0$ and in the long
distance limit $b_{2}(\zeta) = 0$. We note that, in addition to
the conditions specified above, the spatial variation of the
amplitudes should satisfy the adiabaticity condition Eq.\
(\ref{adiab}), as only in this case the solutions (\ref{v1}) and
(\ref{v2}) are valid.

\begin{figure}[h]
\centerline{\scalebox{.75}{\includegraphics{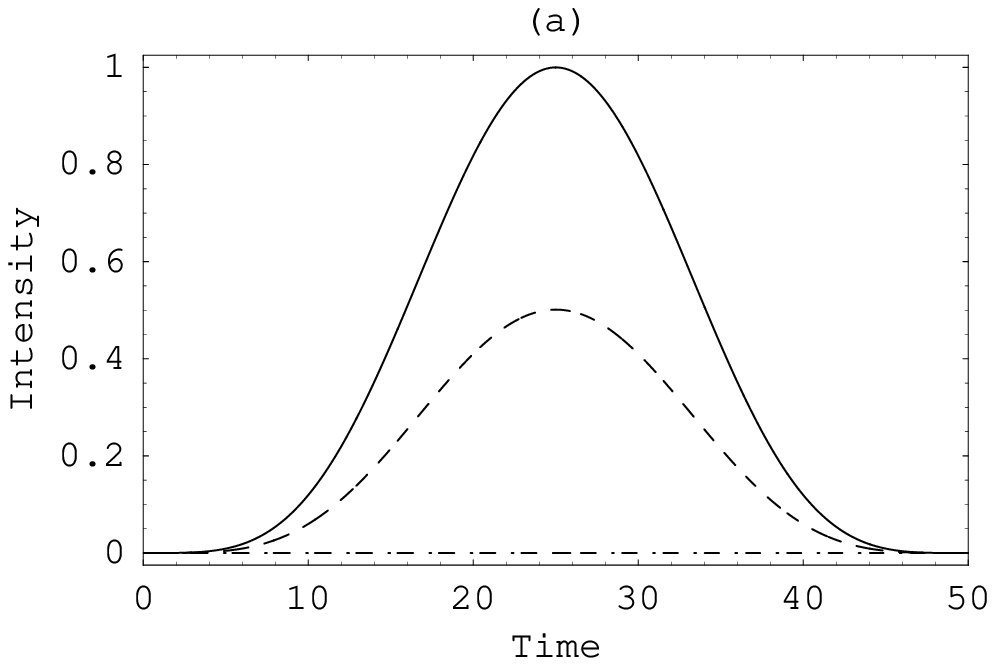}}}

\centerline{\scalebox{.75}{\includegraphics{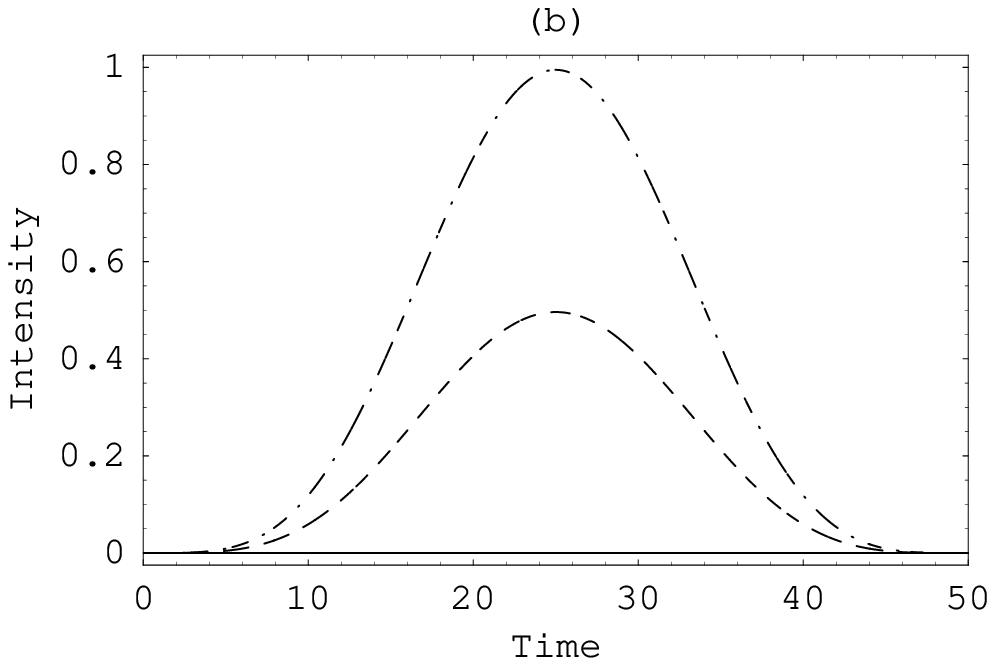}}}

\centerline{\scalebox{.75}{\includegraphics{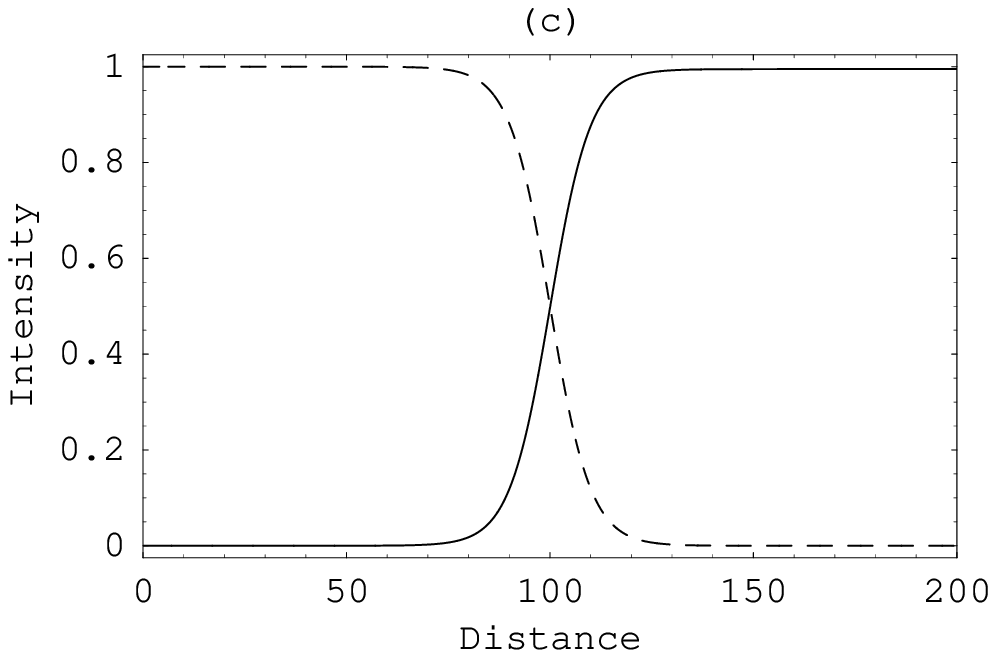}}}

\caption{Paspalakis and Kis}
\label{fig2}
\end{figure}

The process of parametric generation in our system is illustrated in Fig.\
\ref{fig2}, where the spatio--temporal evolution of the normalized intensities
of the laser pulses is shown. The results have been obtained from a numerical
solution of Eq. (\ref{propag}).  The initial spatial distributions are chosen
as
\begin{subequations}
\begin{eqnarray}
  b_{1}(\zeta) &=&
  \sqrt{\frac{1}{1+e^{-(\zeta-\zeta_{0})/\bar{\zeta}}}}\,, \\
  b_{2}(\zeta) &=&
  \sqrt{\frac{1}{1+e^{(\zeta-\zeta_{0})/\bar{\zeta}}}}\,,
\end{eqnarray}
\end{subequations}
and the incoming pulse has a sin-squared shape. It is clear that
the incoming laser pulse is completely converted to a new laser
pulse. The accuracy of the approximations that lead to Eq.\
(\ref{propag}) has been assessed by comparing with the numerical
solution of Eqs.\ (\ref{final1}) and (\ref{v0}). The agreement
between the two results is very good, implying that our
approximations are valid for the chosen parameter set.

In this paper, we have investigated resonant nonlinear optical
processes that occur when a medium is initially prepared in a
spatially dependent coherent superposition. In particular, for a
$\Lambda$-type system we have demonstrated that by proper choice
of the spatially dependent initial excitation complete nonlinear
conversion between two laser pulses is possible. An interesting
application of the present scheme is a coherent polarization
rotator.  Let us consider a $\Lambda$ scheme with lower states
$J_g=1$, $M_g= \pm 1$ and excited state $J_e=0$.  Then, if the
injected field is left-hand circularly polarized, in the medium
it will be transferred smoothly to the other mode and a right-hand
circularly polarized light will exit the medium.

We note before closing that the process studied in this work is
the spatial analog of STIRAP. That is, by using STIRAP in a
$\Lambda$-type system complete population transfer from one state
to another state is possible by proper choice of the Rabi
frequencies.  In our method complete conversion from one laser
pulse to another laser pulse is possible by proper choice of the
initial spatial excitation of the medium.  Both methods are based
on a counter-intuitive excitation and on the adiabatic evolution
of the system.

E.P. would like to thank N.J. Kylstra for assistance in the numerical
simulations. This work was supported by the European Union Research
and Training Network COCOMO, contract HPRN-CT-1999-00129. Z.K.
acknowledges the support of the J\'anos Bolyai program of the
Hungarian Academy of Sciences.

\newpage

\section{List of Figure Captions}

\noindent Fig.1. Schematic diagram of the system studied. The
excited state $|0\rangle$ is coupled by two coherent laser pulses
to the lower states $|1\rangle$ and $|2\rangle$. In the
calculations presented in Fig.\ 2 $\Omega_{1}(\zeta,\tau)$ is the
Rabi frequency of the incoming pulse and $\Omega_{2}(\zeta,\tau)$
is the Rabi frequency of the generated pulse.

\vspace*{1.cm}

\noindent Fig.2. The normalized field intensities
$|\Omega_{1}(\zeta,\tau)|^2/|\Omega_{1}|^2$ for (a), and
$|\Omega_{2}(\zeta,\tau)|^2/|\Omega_{1}|^2$ for (b) as a function
of $\tau$ for different values of $\zeta$, with $\zeta=0$ (solid
curves), $\zeta = 100$ (dashed curves) and $\zeta = 200$
(dot-dashed curves). In figure (c), we present the maximum of the
normalized field intensities as a function of $\zeta$ for the
incoming field (dashed curve) and the generated field (solid
curve). The incident pulse is $\Omega(\tau)$ = $\Omega_1
\sin^{2}(\tau \pi/\tau_{p})$, with $0 \le \tau \le \tau_{p}$. The
parameters used in the calculations are $a_{1}$ = $a_{2}$ =
$1000$, $\Omega_1 = 0.01$, $\tau_{p} = 50$, $\delta = 0$, $\gamma
= 100$, $\zeta_{0}=100$ and $\bar{\zeta} = 5$.  All quantities
are in arbitrary units.

\newpage


\begin{thebibliography}{99}
\bibitem{note}
Usually adiabatic evolution refers to temporal evolution. Here,
we will use it for spatial evolution.
\end{thebibliography}
\end{document}